\documentstyle[preprint,aps,floats]{revtex}
\tightenlines
\input psfig.tex
\begin{document}
\def\ba{\begin{eqnarray}}
\def\ea{\end{eqnarray}}
\def\be{\begin{equation}}
\def\ee{\end{equation}}
\def\({\left(}
\def\){\right)}
\def\[{\left[}
\def\]{\right]}
\def\lagrange {{\cal L}}
\def\del {\nabla}
\def\d {\partial}
\def\Tr{{\rm Tr}}
\def\half{{1\over 2}}
\def\fourth{{1\over 8}}
\def\bibi{\bibitem}
\def\S{{\cal S}}
\def\xx{\mbox{\boldmath $x$}}
\newcommand{\labeq}[1] {\label{eq:#1}}
\newcommand{\eqn}[1] {(\ref{eq:#1})}
\newcommand{\labfig}[1] {\label{fig:#1}}
\newcommand{\fig}[1] {\ref{fig:#1}}
\def\gsim{ \lower .75ex \hbox{$\sim$} \llap{\raise .27ex \hbox{$>$}} }
\def\lsim{ \lower .75ex \hbox{$\sim$} \llap{\raise .27ex \hbox{$<$}} }
\newcommand\bigdot[1] {\stackrel{\mbox{{\huge .}}}{#1}}
\newcommand\bigddot[1] {\stackrel{\mbox{{\huge ..}}}{#1}}
\title{Open Inflation Without False Vacua
} 
\author{S.W. Hawking\thanks{email:S.W.Hawking@damtp.cam.ac.uk} 
and Neil
Turok\thanks{email:N.G.Turok@damtp.cam.ac.uk}}
\address{
DAMTP, Silver St, Cambridge, CB3 9EW, U.K.}
\date\today 
\maketitle

\begin{abstract}
We show that within the framework of a definite proposal for
the initial conditions for the universe, the Hartle-Hawking `no boundary' 
proposal, open inflation is generic and does not require any special
properties of the inflaton potential. In the simplest inflationary 
models, the semiclassical
approximation to the Euclidean
path integral and a minimal anthropic condition lead to
$\Omega_0\approx 0.01$. This number may be increased in models
with more fields or extra dimensions.
\end{abstract}
\vskip .2in

\section{Introduction}

The inflationary universe scenario 
provides an appealing explanation 
for the size, flatness and smoothness of the present universe, 
as well as a mechanism for the origin of fluctuations. 
But whether inflation actually occurs within a given 
inflationary model is known to 
depend very strongly on the pre-inflationary
initial conditions. In the absence of a measure on the
set of initial conditions inflationary theory inevitably
rests on ill-defined foundations.
One such measure is provided by
continuing the path integral to imaginary time
and demanding that the Euclidean four manifold so obtained be compact
\cite{HH}. This is the Hartle-Hawking `no boundary' proposal.
In this Letter we show that the no boundary
prescription, coupled to a minimal anthropic condition, actually 
predicts  open inflationary universes for generic scalar 
potentials. The simplest inflationary potentials with 
a minimal anthropic requirement favour values of 
$\Omega_0 \sim 0.01$, but generalisations including extra fields
favour more reasonable values.
At the very least these calculations demonstrate that the
measure for the pre-inflationary initial conditions {\it does} matter. 
More importantly, we believe the implication is that inflation itself is
now seen to be perfectly compatible with
an open universe.

Until recently it was believed that all inflationary models  
predicted $\Omega_0=1$ 
to high accuracy.
This view was overturned by the discovery that 
a special class of inflaton potentials produce nearly homogeneous
open universes with interesting values of $\Omega_0<1$ today \cite{gott}, 
\cite{bucher}.
The potentials were required to have a metastable minimum (a `false vacuum') 
followed by a gently sloping region allowing 
slow roll inflation.
The idea was that the inflaton could become
trapped in the 
`false vacuum', driving a period of inflation and creating a 
near-perfect De Sitter space with minimal quantum fluctuations. 
The field would then quantum tunnel,
nucleating bubbles within which it would roll slowly down to the true
minimum. 
The key observation, due to Coleman and De Luccia \cite{coleman} 
is that the interior of such a bubble is actually
an infinite open universe. 
By adjusting the duration of the slow roll epoch one can arrange that
the spatial curvature today is of order the Hubble radius \cite{bucher}.

All inflationary models must be fine tuned to keep the 
quantum fluctuations small. This requires that the potentials
be very flat. In open inflation 
this must be reconciled with  
the requirement that
the potential   
have a false vacuum. Furthermore, a 
classical bubble solution of the Coleman De Luccia 
form 
only exists
if the mass of the scalar
field in the false vacuum is large, so that the bubble `fits inside' the 
De Sitter Hubble radius. 
Taken together these requirements meant that the scalar potentials needed 
for open inflation were very contrived for single field models.
Two-field models were proposed, but even these required a false vacuum 
\cite{lindeopen}) and the pre-bubble initial conditions were
imposed essentially by hand.

Within the  Hartle-Hawking framework, the period of `false vacuum' inflation
is no longer required.
The quantum fluctuations are computed by
continuing the field and metric perturbation modes from the 
Euclidean region
where they are governed by a positive definite measure.
The Hartle-Hawking prescription in effect 
starts 
the universe in a state where the fluctuations 
are at a minimal level in the first place.

\section{Instantons}

We consider the path integral for 
Einstein gravity coupled to a scalar field $\phi$,
with potential $V(\phi)$, which we assume has a true minimum with 
$V=0$. As usual, we approximate the path integral by 
seeking saddle points i.e. solutions of the classical equations of motion,
and expanding
about them to determine the fluctuation measure.
We begin with the Euclidean instanton. 
If $V(\phi)$ has a stationary 
point at some nonzero value then
there is an $O(5)$ invariant solution
where 
$\phi$ is constant and the 
Euclidean manifold is a four sphere.
We shall be interested in more general solutions 
possessing only $O(4)$ invariance.
The metric takes the form
\be
ds^2= d\sigma^2 +b^2(\sigma)d \Omega_3^2 =d\sigma^2 +b^2(\sigma)(d \psi^2+{\rm sin}^2 (\psi) d \Omega_2^2)
\labeq{emetric}
\ee
with $b(\sigma)$ the radius of the $S^3$ `latitudes' of the $S^4$. For the
$O(5)$ invariant solution 
$b(\sigma)=H^{-1} {\rm sin}
(H\sigma)$, with $H^2= 8 \pi G V/3$, but in
the general case  $b(\sigma)$
is a deformed version of the sine function. 

\begin{figure}
\centerline{\psfig{file=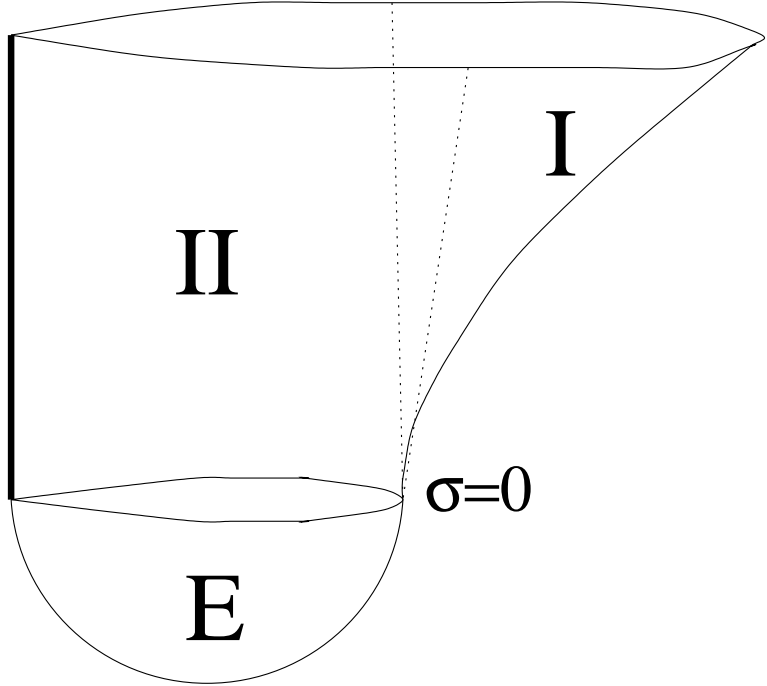,width=4.in}}
\caption{Global structure of the open instanton and its continuation.
The Euclidean region E is half of a deformed four sphere. It continues into
a De Sitter like region II, and thence into an open inflating universe, region I. 
The dotted lines show the null surface (the `bubble wall') emanating from the 
point $\sigma=0$ on the instanton. The heavy line shows the singularity discussed in the
text.
}
\labfig{sketch}
\end{figure}

Solutions possessing only $O(4)$ invariance are
naturally continued to an open
universe as follows (Figure 1).
First we  continue from Euclidean to Lorentzian space.
To obtain a real Lorentzian metric we must continue on a three 
surface where the metric is stationary (more properly, where the
second fundamental form vanishes). One obtains an open universe by
continuing $\psi$,
so that $\psi$ runs from $0$ to $\pi/2$
in the Euclidean region and then in the imaginary direction in the Lorentzian
region.
Setting 
$\psi =\pi/2 +i\tau$ we obtain 
\be
ds^2= d\sigma^2 +b^2(\sigma)(-d\tau^2+{\rm cosh}^2 (\tau) d \Omega_2^2).
\labeq{twometric}
\ee
which is a spatially inhomogeneous De Sitter-like metric. 
This metric describes region II of the solution, the exterior of the
inflating bubble. The radius $b(\sigma)$ vanishes at two values of 
$\sigma$. Near the the first, which we shall call
$\sigma=0$, $b(\sigma)$ vanishes linearly with $\sigma$.
The metric has a unique continuation through the null surface defined by
$\sigma=0$. One sets 
$\sigma=it$ and $\tau=i\pi/2+\chi$ giving
\be
ds^2= -dt^2 +a^2(t)(d \chi^2+{\rm sinh}^2 (\chi) d \Omega_2^2)
\labeq{onemetric}
\ee
where $a(t)=-ib(it)$. This is an expanding open universe describing region
I of the solution.

There is another inequivalent
continuation from the Euclidean instanton which 
produces a closed universe. This is obtained by continuing the 
coordinate $\sigma$ in the imaginary direction beyond the value
$\sigma_{max}$ at which the radius $b(\sigma)$ is greatest.
So $\sigma$ runs from 0 to $\sigma_{max}$ in the 
Euclidean region, and  
$\sigma=\sigma_{max}+iT$ in the Lorentzian region. 
The latter is a
De Sitter-like space with 
homogeneous but time dependent spatial sections:
\be
ds^2= -dT^2 +b^2(T)(d \psi^2+{\rm sin}^2 (\psi) d \Omega_2^2).
\labeq{cmetric}
\ee
We shall return to this solution later  - it describes
a closed inflating universe. 

Now let us discuss the properties of the Euclidean 
instanton in more detail. The field $\phi$ and the radius $b$ obey
the field equations
\be
\phi''+3{b'\over b}\phi'=V_{,\phi},\qquad b''= -{8\pi G\over 3} b (
\phi'^2 +V) 
\labeq{field}
\ee
where primes denote derivatives with respect to $\sigma$.
According to the first equation, $\phi$ rolls 
in the upside down potential $-V$. The point $\sigma=0$ is assumed 
to be a nonsingular point so the manifold 
looks locally like $R^4$ in spherical polar coordinates. 
This requires that $b(\sigma) \sim \sigma$ 
at small $\sigma$. 
The field takes the value $\phi_0$ at $\sigma=0$. We
assume the potential has 
a nonzero slope at this field value $V_{,\phi}(\phi_0)\neq 0$ 
(otherwise we would obtain the $O(5)$ invariant instanton).
Analyticity and $O(4)$ invariance imply that 
$\phi'(0) = 0$.
Following the solutions forward in $\sigma$, $b(\sigma)$ decelerates
and its velocity 
$b'(\sigma)$ changes sign. Thereafter $b(\sigma)$ is driven to zero, at a point
we call
$\sigma_{f}$. The field $\phi$ on the other hand is driven  {\it} up 
the potential by the forcing term, initially with damping but after the
sign change in $b'(\sigma)$ with antidamping. The antidamping diverges as 
we approach $\sigma_{f}$, and $\phi'(\sigma)$ goes to infinity there. 
As we approach $\sigma_f$ the potential terms become irrelevant in the
field equations: the first
equation then implies that $\phi' \propto b^{-3}$ and the second
yields $b\propto (\sigma_{f}-\sigma)^{1\over 3}$. Thus
$\phi' \propto (\sigma_{f}-\sigma)^{-1}$ and 
$\phi$ diverges logarithmically 
as we approach the singularity. The above behaviour is true for 
any $\phi_0$ if the potential increases monotonically away from 
the true minimum.
If there are additional extrema
it is possible for the driving term $V_{,\phi}$ to change sign and,
if it is large enough to counteract the antidamping term, to
actually stop the motion of $\phi$.
The Coleman-De Luccia instanton
is obtained only for potentials where this is possible 
(see e.g. \cite{steinh}). It 
occurs when the value of $\phi_0$ is chosen so that $\phi'$ 
returns to zero 
precisely at $\sigma_{f}$. In that case,
{\it} both ends of the solution are nonsingular and a continuation into
a third Lorentzian region becomes possible. 
The Coleman-De Luccia instanton was employed in 
previous versions of open inflation
because it is unique and nonsingular, in analogy with 
tunnelling solutions in Minkowski space. 
But De Sitter space is quite different from Minkowski space, possessing
finite closed spatial sections, and the question of which instantons
are allowed needs to be separately examined.

The primary criterion for deciding 
whether an
instanton solution is physically allowed is to 
compute the Euclidean action $S_{E}$. 
The wavefunction for the system is in the leading approximation
proportional to $e^{-S_E}$ so configurations of infinite action are
suppressed. 
The Euclidean action is given by 
\be
S_{E}= \int d^4x 
\sqrt{g}\left[-{R \over 16 \pi G}+{1\over 2} (\partial \phi)^2 +V\right].
\labeq{eact}
\ee
But in four dimensions 
the trace of the Einstein  equation reads 
$R=8 \pi G ((\partial \phi)^2+4V(\phi))$ and so the action is just
\be
S_{E}=-\int d^4x \sqrt{g} V =  -\pi^2 \int d\sigma b^3(\sigma) V(\phi).
\labeq{evact}
\ee
where we have integrated over half of the $S^3$. Note that 
the action is {\it negative}, a result of the well known 
lack of positivity 
of the Euclidean gravitational action. The surprising thing however 
is that even for our singular instantons, at
the singularity
$V$ diverges only logarithmically. The volume measure $b(\sigma)^3$ 
vanishes linearly with
$(\sigma_f-\sigma)$ so the Euclidean action 
is perfectly convergent. If one examines more closely how this 
result emerges, one finds that the scalar field 
part of the action diverges logarithmically (since 
$\phi'$ diverges linearly) but this divergence is precisely cancelled
by an opposite divergence in the gravitational action. There are 
two key differences between the present calculation and that for tunnelling
in Minkowski space. First, the instanton is spatially finite and this
cuts off the divergence associated with the field not tending to 
a minimum of the potwntial. Second, 
the gravitational action is not positive and is thus able to
cancel a divergence in the scalar field action. These two differences
have the remarkable consequence that unlike the situation in 
Minkowski space, there is a 
a one parameter
family of allowed instanton solutions. 

Let us now comment 
on the singularity at $\sigma_f$, which is timelike. 
Timelike spacetime singularities are not necessarily fatal in 
semiclassical descriptions of quantum physics, as the example of
the hydrogen atom teaches us. Generic particle trajectories
`miss' the singularity, and quantum fluctuations may be enough to
smooth out its effect. 
In the present case we shall see that
the singularity is 
mild enough for the quantum 
field fluctuations to be well defined.
The field and metric
fluctuations are defined by continuation from
the Euclidean region, singular only at a point on its edge. 
The mode functions for the field fluctuations are most easily
studied by changing coordinates from $\sigma$ to 
the  conformal
coordinate
$X= \int^{\sigma_f}_{\sigma}
d\sigma/b(\sigma)$. Because the integral converges at
$\sigma_f$, the range of $X$ is bounded below by zero. 
After a rescaling
$\phi = \chi/b$, the field modes
obey a Schrodinger-like 
equation with a potential given by 
$b^{-1} (d^2 b /dX^2) - V_{,\phi \phi} b^2\sim -{1\over 4} X^{-2}$
at small $X$. This divergence is precisely critical - 
for more negative coefficients an inverse square potential has a 
continuum of
negative energy states and the quantum mechanics is pathological. But for
$-{1\over 4} X^{-2}$
there is a 
positive continuum and a well defined complete set of 
modes. 
The causal structure of region II is easily seen  in the same
conformal coordinates.
Near the singularity the spatial metric of region II
is conformal to a tube $R^+ \times S^2$. The singularity
is a world line corresponding to 
the end of the tube. 

As mentioned above, there is another instanton describing a closed 
inflationary universe where one continues $\sigma$ in the imaginary direction
from $\sigma_{max}$. The action of this
instanton is given by twice the expression 
(\ref{eq:evact}) but with the integral taken only over
the interval $[0,\sigma_{max}]$. The functions $b(\sigma)$ and $\phi(\sigma)$ 
are also somewhat different - analyticity still 
implies that $\phi'(0)$ must be zero, but since the potential has
a nonzero slope,  
the velocity $\phi'$ 
is nonzero on the matching surface. This leads to odd terms in the 
Taylor expansion for $\phi$ around $\sigma=0$, so 
$\phi(T)$ is complex in the 
Lorentzian region. One would like the solution for $\phi$ 
be real at late times. This is impossible to arrange exactly, but
one can add a small imaginary part to $\phi_0$ in such a way that
the imaginary part of $\phi$ is in the pure 
decaying mode during inflation.
Then both the
field and metric are real to exponential accuracy at late times.

The potentials of interest are those
whose slope 
is sufficiently shallow to allow many inflationary efoldings.
We have numerically computed 
the action as a function of the 
parameter $\phi_0$ for various scalar potentials.
In the regime where the number of efoldings is large, the result is 
very simple - to a good approximation one has
$\phi(\sigma)\approx \phi_0$ and 
$b(\sigma) \approx H^{-1}{\rm sin} H\sigma$ over most of the range of $\sigma$,
where
$H^2=8 \pi G V(\phi_0)/3$. The Euclidean action is
then just 
\be
S_{E} \approx - {12 \pi^2 M_{Pl}^4 \over V(\phi_0)},
\labeq{estact}
\ee
in both open and closed cases,
where the reduced Planck mass $M_{Pl} = (8 \pi G)^{-{1\over 2}}$.

\section{ The Value of $\Omega_0$}

The value of the density parameter today, $\Omega_0$, is determined
by the number of inflationary efoldings. On the relevant matching
surface the value of $\Omega$ is zero in the open case, infinity 
in the closed case. It 
approaches unity as $\Omega^{-1}-1 \propto a^{-2}$ during inflation.
After 
reheating it deviates from unity 
as $\Omega^{-1}-1 \propto a^{2}$ in the radiation era 
and $\Omega^{-1}-1 \propto a$
in the matter era. 

Putting this together, and assuming instantaneous reheating, 
one finds \cite{bucher} that 
\be
\Omega_0 \approx {1\over 1 \pm  {\cal A} e^{-2 N(\phi_0)}},  \qquad {\cal A}
\approx 4 \left( {T_{reheat}\over T_{eq}}\right)^2 {T_{eq}\over T_0}
\labeq{omega}
\ee
where the $+$ and $-$ refer to
the open and closed cases respectively. The temperature today is $T_0$,
that at matter-radiation equality is $T_{eq}$. We 
assume that $T_{eq}>T_0$, otherwise one should set $T_{eq}=T_0$. 
The constant ${\cal A}$ depends on the reheating temperature - it
ranges between $10^{25}$ and $10^{50}$ for reheating to the electroweak 
and GUT scales respectively.

The number of inflationary efoldings is given in the slow roll 
approximation by 
\be
N(\phi_0) \approx \int^{\phi_0} d \phi {V(\phi)\over V_{,\phi}(\phi) M_{Pl}^2}
\labeq{nefo}
\ee
where the lower limit is the value of $\phi$ where the slow roll condition
is first violated. For example, for a quadratic potential
$N \sim (\phi_0/ 2 M_{Pl})^2.$ For small $\phi_0$ there are few efoldings
and $\Omega_0$ is very small in the open case, or the universe 
collapsed before $T_0$ in the closed case. For large $N$, 
$\Omega_0$ is very close to unity. But for $N$ in the range 
${1\over 2}{\rm log} {\cal A}
\pm 1$, which is $30 \pm 1$ or $60 \pm 1$ for reheating to the 
electroweak or GUT scales respectively, we have $0.1 <\Omega_0 <0.9$.
So some tuning of $\phi_0$ is 
required to obtain interesting values for $\Omega_0<1$ today, 
but it is only logarithmic and therefore quite mild \cite{bucher}.

The formula (\ref{eq:omega}) involves several unknown parameters, 
and depending on the context one has to decide which of them to keep fixed.
The Einstein equations for matter, radiation and curvature allow
three independent constants, which may be taken as $H_0$, $\Omega_0$ and $T_0$. 
The temperature at matter-radiation equality $T_{eq}$ is not independent
since it is determined by the matter density today, 
fixed by $\Omega_0$ and $H_0$, and the radiation density today,
fixed by $T_0$. In principle $T_{eq}$ it is 
determined in terms of
the fundamental Lagrangian just as the temperature at decoupling is, 
but since we do not know the Lagrangian it is better to
eliminate  $T_{eq}$
using 
$T_{eq}= 2.4 \times 10^4 \Omega_0 h^2 T_0$. 
This introduces $\Omega_0$ dependence into the right hand side of 
(\ref{eq:omega}), so one solves to obtain
\be
\Omega_0 \approx 1 \mp  {\cal A'} e^{-2 N(\phi_0)},  \qquad {\cal A'}
\approx 4 \left( {T_{reheat}^2\over 2.4\times 10^4 h^2 T_{0}^2}\right). 
\labeq{omegap}
\ee
(For the open case if $\Omega_0 < (2.4 \times 10^4 h^2)^{-1}$
 and $T_0>T_{eq}$ one should
use 
(\ref{eq:omega}) with
$T_{eq}$ replaced by $T_0$). The formula (\ref{eq:omegap}) gives 
us $\Omega_0$ in terms of the presently observed parameters
$T_0$ 
and $H_0$, plus the inflationary parameters namely the 
initial field $\phi_0$ and the reheat 
temperature $T_{reheat}$. 

Let us summarise the argument so far. We have constructed families of 
complete background solutions describing open 
and closed inflationary universes for essentially any
inflaton potential. These solutions solve the standard inflationary 
conundrums, since exponentially large, homogeneous universes 
are obtained from initial data specified within a single Hubble
volume. Each also has a well defined spectrum of fluctuations
obtained by analytic continuation from the Euclidean 
region. It is worthwhile to explore how well these 
solutions, and their associated perturbations, match the
observed universe. We shall do so in future work. 

More ambitiously, one can also attempt to understand
the theoretical probability distribution for $\Omega_0$,
and it is to this that we turn next.

\section{ Anthropic Estimate of $\Omega_0$}

The {\it a priori} probability for 
a universe to have given value of $\Omega_0$ is
proportional the square of the wavefunction, given 
in the leading semiclassical approximation by $\propto e^{-2S_E}$.
We will work in some fixed 
theory in which $T_{reheat}$ is determined by the Lagrangian.
The initial field 
$\phi_0$ is however still a 
a free parameter labelling the relevant instanton. We consider
a generic inflationary potential which increases away from 
zero. 
Both closed and 
open solutions exist for arbitrarily large 
$\phi_0$, so at
least for suitably flat potentials essentially
all possible values of $\Omega_0$ are allowed. There are also  
closed solutions  where
${\cal A} e^{-2 N}>1$, in which the universe turns round and 
recollapses before ever reaching the present temperature
$T_0$.

The Euclidean action (\ref{eq:estact}) is  
typically {\it huge} - and in the simplest theories 
is 
likely to be the dominant factor in the probability
distribution $P(\Omega_0)$.
The most favoured universes are those with the smallest initial field $\phi_0$: 
these universes are either essentially empty at $T_0$ in the open case, 
or recollapsed long before $T_0$ in the closed case.
These universes are quite different from our own, and one might be tempted to
discard the theory.
Before doing so, we might remind the reader 
that all other versions of inflation fail {\it just as badly} in this regard
- they are just less mathematically explicit about the problem.
According to the heuristic picture of chaotic inflation for example, 
an exponentially large fraction of the universe is still inflating, 
and we certainly do not inhabit a typical region.
So as in that case (and with some reluctance!)
we shall be forced to make an anthropic argument.

If one knew the precise
conditions required for the formation of observers it would 
be reasonable to restrict attention
to the subset of universes containing them.
The problem is that we do not. The best we can do is to make 
a {\it guess} based on our poor knowledge of the requirements for
the formation of life, namely the production of heavy elements in stars 
and a reasonably long time span to allow evolution to take place. 
Such an invocation of the anthropic principle represents
a retreat for theory - we give up on the goal of 
explaining all the properties of the universe by using some (our existence)
to 
constrain others (e.g. $\Omega_0$). However we don't think it is 
completely unreasonable, and it may (unfortunately!) turn out to be essential.
An alternative attitude is to seek a future theoretical development 
that will fix the parameter $\phi_0$
and the problem of its probability distribution. Both avenues are 
in our view  worth pursuing.

The anthropic condition is naturally implemented within a Bayesian framework
where one regards the wavefunction as giving the prior probability
for $\Omega_0$, and then computes the posterior probability for 
$\Omega_0$ given the fact that our galaxy formed. So one writes
\be
{\cal P} (\Omega_0|{\rm gal}) \propto {\cal P} ({\rm gal}|\Omega_0) 
{\cal P} (\Omega_0) \propto 
 {\rm exp}\left(-{\delta_c^2 \over 2 \sigma_{gal}^2} -2S_E(\phi_0)\right)
\labeq{prob}
\ee
where the first factor represents the probability that the galaxy-mass
region in our vicinity underwent gravitational collapse, for given $\Omega_0$. 
The rms contrast of the linear density field smoothed on the galaxy mass scale
today is $\sigma_{gal}$, and $\delta_c \approx 1$
is the threshold set on the
linear perturbation amplitude by the requirement that gravitational collapse
occurs. 
We have only included the leading exponential terms in
(\ref{eq:prob}), and  
have assumed Gaussian perturbations as predicted by
the simplest inflationary models.

The rms contrast in the density field today $\sigma_{gal}$ is 
given by the perturbation 
amplitude at Hubble radius crossing for the galaxy scale $\Delta(\phi_{gal})$ 
multiplied by
the growth factor $G(\Omega_0)$. The latter is strongly dependent
on 
$\Omega_0$
both through the redshift of matter-radiation 
equality
and the loss of growth at late times in a low density universe
\cite{Peebles}. Roughly one has $G(\Omega_0) \sim 2.4 \times 10^{4} h^2
\Omega_0^2 \sim 10^4 \Omega_0^2$ for $h=0.65$. 
In the slow roll approximation the linear perturbation amplitude
at horizon crossing is
\be
\Delta^2(\phi)  \equiv {V^3 \over M_{Pl}^6 V_{,\phi}^2}.
\labeq{pert}
\ee

At this point it is interesting to compare and contrast the open and
closed inflationary continuations. If we fix $\phi_0$, the 
Euclidean actions and therefore the prior probabilities are
very similar. From (\ref{eq:omegap}) one sees that for fixed 
$H_0$ and $T_0$, 
an open universe with density parameter $\Omega_0$ is as likely 
{\it a priori} as a closed universe with density parameter $2-\Omega_0$. Of 
course the two universes are very different. 
The first difference is that the closed universe is considerably
younger - for $h=0.65$ the open universe is 15 Gyr old, the closed one is
8 Gyr old. The second and most striking difference is that 
the 
open universe
is spatially infinite whereas the closed universe is finite. 
If one accepted the arguments of some other authors \cite{VWin,weinberg}
that the number of observers is the determining factor, one would conclude that
open inflation was infinitely more probable because it would produce an
infinite number of galaxies. However we do not agree with this line of 
reasoning because it would be like arguing that we are more likely to
be ants because
there are more ants than people! For this reason we prefer to use 
Bayesian statistics and consider the probability of forming a galaxy 
at fixed $H_0$ and $T_0$ rather than the total number of galaxies. 

In the open case, the galaxy formation probability produces a peak in 
the posterior probability for $\Omega_0$.
At very low
$\Omega_0$ the growth factor is so small that 
galaxies become exponentially rare. 
From (\ref{eq:omegap}) 
\be
{d\Omega_0 \over d \phi_0}=  {2 V\over M_{Pl}^2 V_{,\phi}} (1-\Omega_0), 
\labeq{expopd}
\ee
and it follows that the most likely value for $\Omega_0$ is
given by 
\be
\Omega_0 \approx 0.01 
\left({\Delta^2(\phi_{gal}) \over \Delta^2(\phi_0)} 
\right)^{1\over 5}.
\labeq{peaka}
\ee
The simplest inflationary models are close to being 
scale invariant, so the latter factor is close to unity. 
The result, $\Omega\approx 0.01$,
is interestingly close to the baryon density 
required for primordial nucleosynthesis, but
too low to be compatible with current
observations. 

According to these arguments 
the most probable open universe is one where matter-radiation
equality happened at a redshift of 100, well after decoupling. 
The horizon scale at that epoch is $\sim 2500 h^{-1}$ Mpc (for $h=0.65$)
and for a pure baryonic universe the power spectrum for matter
perturbations would be scale invariant from that scale down to
the Silk damping scale, an order of magnitude smaller. The nonlinear
collapsed region around us would be somewhere between these
scales in size. It would be an isolated, many sigma high density peak 
surrounded by a very low density universe.  Interestingly,
the value of $\Omega_0$ we would measure would be much higher than
the global average. However even though such a region would be 
large, it is hard to see how the universe would appear as 
isotropic as it does to us (in the distribution of
radio galaxies and X rays for example)
unless we lived in the centre of the collapsed region, and it
was nearly spherical. 

In the closed case, the prior probability distribution favours 
universes which recollapsed before the temperature ever reached $T_0$. 
If we fix $T_0$ and $H_0$ (i.e. demanding the universe be expanding)
a peak in the posterior probability for
$\Omega_0$ is produced by imposing the anthropic 
condition that the universe
should be old enough to allow the evolution of life, say 5 billion years. 
For a Hubble constant $h=0.65$, this requires that $\Omega_0 < 10$,
and the peak in the posterior probability would be at $\Omega_0=10$. 
If we raised this age requirement to 10 billion years, the most 
likely value for $\Omega_0$ would be just above unity. 
The most likely closed universe would be more probable than the 
most likely open universe in the first case, but less probable 
in the second. 

Even though these most likely universes (i.e. very closed
or very open) are probably not an acceptable fit
to our own, we nevertheless
find it striking that such 
simple 
arguments lead to 
a value of 
$\Omega_0$ not very far from the real one. The simple
inflationary models we have discussed here are certainly not
final theories of quantum gravity, and it is quite possible that
a more complete theory would lead to a modified distribution 
for $\phi_0$ giving a more acceptable values of $\Omega_0$. 
In particular it seems possible that the prior distribution 
for $\Omega_0$ would favour values closer to unity while
disfavouring intermediate values, but one would still need
to invoke anthropic arguments to exclude very high or very low values.

One possible mechanism for increasing the probability of a high initial
field value $\phi_0$ and therefore a value for $\Omega_0$ nearer 
unity might just be phase space. In a realistic theory, with
many more fields, there are an infinite number of instanton solutions
of the type we have discussed. Each starts at some point in field
space, with the fields rolling up the potential in the instanton and
down the potential in the open universe. If we assume one field $\phi$
provides most of the inflation, it is possible that
as $\phi_0$ increases, the other fields it couples to become massless.
This would increase the phase space available at given Euclidean
action. For example, $\phi$ could be the scalar field parametrising the 
radius of an extra dimension: in this case the 
radius $R$ would be proportional to $e^{(\phi/M_{Pl})}$. Then $\phi$ getting
large would mean that the tower of Kaluza Klein modes became exponentially
light and there would be a corresponding exponential growth in 
the phase space available at fixed Euclidean action. This exponential
growth in phase space would cease when the extra dimension became so large
that the extra dimensional gauge coupling became of order unity, and one entered
the strong coupling regime.

In summary, we have proposed a new framework for inflation 
in which values for the density parameter $\Omega_0 \neq 1$ are
allowed for generic inflaton potentials, whilst retaining the
usual successes of inflation including a predictive pattern of
density perturbations. 
The generic prediction of this framework
is a very open or very closed universe, 
but it is possible that including other fields
and extra dimensions could result in more acceptable values
of $\Omega_0$ closer to unity.

\medskip
\centerline{\bf Acknowledgements}

We would like to thank R. Crittenden and S. Gratton for helpful 
discussions. 


\end{document}